\documentclass{aa}  

\usepackage{amssymb}
\usepackage{amsmath}
\usepackage{graphicx}
\usepackage{txfonts}
\usepackage{booktabs}
\usepackage{textcomp}
\usepackage{color}
\usepackage{lscape}
\usepackage{longtable}
\usepackage{xcolor}
\bibpunct{(}{)}{;}{a}{}{,}

\begin{document}

  \title{A search for prompt gamma--ray counterparts to fast radio bursts in the {\em Insight}-HXMT data}
\titlerunning{A search for FRB counterparts in {\em Insight}-HXMT data}
\authorrunning{C.~Guidorzi et al.}
 
   \author{C.~Guidorzi\thanks{guidorzi@fe.infn.it}\inst{1,2,3} \and M.~Marongiu\inst{1} \and R.~Martone\inst{1} \and L.~Nicastro\inst{3} \and S.L.~Xiong\inst{4} \and J.Y.~Liao\inst{4} \and G.~Li\inst{4} \and S.N.~Zhang\inst{4,5} \and L.~Amati\inst{3}
% ...
}
\institute{Department of Physics and Earth Science, University of Ferrara, via Saragat 1, I--44122, Ferrara, Italy
\and INFN -- Sezione di Ferrara, via Saragat 1, I--44122, Ferrara, Italy
\and INAF -- Osservatorio di Astrofisica e Scienza dello Spazio di Bologna, Via Piero Gobetti 93/3, I-40129 Bologna, Italy
\and Key Laboratory of Particle Astrophysics, Institute of High Energy Physics, Chinese Academy of Sciences, 19B Yuquan Road, Beijing 100049, People’s Republic of China
\and University of Chinese Academy of Sciences, Chinese Academy of Sciences, Beijing 100049, China
\and Department of Astronomy, Beijing Normal University, Beijing 100088, People’s Republic of China
\and Department of Physics, Tsinghua University, Beijing 100084, People’s Republic of China
\and Department of Astronomy, Tsinghua University, Beijing 100084, People’s Republic of China
\and Department of Engineering Physics, Tsinghua University, Beijing 100084, People’s Republic of China
\and Computing Division, Institute of High Energy Physics, Chinese Academy of Sciences, 19B Yuquan Road, Beijing 100049,People’s Republic of China
\and Key Laboratory of Space Astronomy and Technology, National Astronomical Observatories, Chinese Academy of Sciences, Beijing 100012,People’s Republic of China 
\and School of Physics and Optoelectronics, Xiangtan University, Yuhu District, Xiangtan, Hunan, 411105, China
\and College of Physics, Jilin University, No.2699 Qianjin Street, Changchun City, 130012, China
}
\date{}

\abstract
   {No robust detection of prompt electromagnetic counterparts to fast radio bursts (FRBs) has yet been obtained, in spite of several multi-wavelength searches carried out so far. Specifically, X/$\gamma$--ray counterparts are predicted by some models.}
   {We planned on searching for prompt $\gamma$--ray counterparts in the {\em Insight}-Hard X-ray Modulation Telescope ({\em Insight}-HXMT) data, taking advantage of the unique combination of large effective area in the keV--MeV energy range and of sub-ms time resolution.}
   {We selected 39 FRBs that were promptly visible from the High-Energy (HE) instrument aboard {\em Insight}-HXMT. After calculating the expected arrival times at the location of the spacecraft, we searched for a significant excess in both individual and cumulative time profiles over a wide range of time resolutions, from several seconds down to sub-ms scales.  Using the dispersion measures in excess of the Galactic terms, we estimated the upper limits on the redshifts.}
   {No convincing signal was found and for each FRB we constrained the $\gamma$--ray isotropic-equivalent luminosity and the released energy as a function of emission timescale. For the nearest FRB source, the periodic repeater FRB\,180916.J0158+65, we find $L_{\gamma,{\rm iso}}<5.5\times10^{47}$~erg/s over 1~s, whereas $L_{\gamma,{\rm iso}}<10^{49}$--$10^{51}$~erg/s for the bulk of FRBs. The same values scale up by a factor of $\sim100$ for a ms-long emission.}
   {Even on a timescale comparable with that of the radio pulse itself no keV--MeV emission is observed. A systematic association with either long or short GRBs is ruled out with high confidence, except for subluminous events, as is the case for core-collapse of massive stars (long) or binary neutron star mergers (short) viewed off axis. Only giant flares from extra-galactic magnetars at least ten times more energetic than Galactic siblings are ruled out for the nearest FRB.}
\keywords{FRB -- radiation mechanism}
\maketitle
%
\newcommand{\commcg}[1]{\textcolor{red}{Cristiano: #1}}
\newcommand{\commmm}[1]{\textcolor{teal}{Marco: #1}}
\newcommand{\commln}[1]{\textcolor{blue}{Luciano: #1}}
\newcommand{\commla}[1]{\textcolor{orange}{Lorenzo: #1}}
\newcommand{\commrm}[1]{\textcolor{gray}{Renato: #1}}
\newcommand{\commff}[1]{\textcolor{cyan}{Filippo: #1}}
\newcommand{\commmo}[1]{\textcolor{olive}{Mauro: #1}}

\section{Introduction}
Fast radio bursts (FRBs) are ms-long, bright ($\sim$Jy) flashes of unknown extra-galactic origin, that have become the focus of a global scientific community since their discovery \citep{Lorimer07,Thornton13}. Despite the recently booming discovery rate and the rapid succession of new findings, their origin remains mysterious (see \citealt{Katz18rev,CordesChatterjee19,Petroff19_rev} for reviews). The growing sample of repeating sources vs. one-off events \citep{CHIME19c,Kumar19,Fonseca20} was somehow expected from considerations based on the relative volumetric rate compared with other cataclysmic sources that could be possibly associated, such as gamma--ray burts (GRBs) or some kinds of supernovae \citep{Ravi19c}. The variety of the few host galaxies so far identified \citep{Tendulkar17,Bannister19,Prochaska19,Ravi19b,Marcote20}, for both repeaters and one-off sources, adds to the enigma of the progenitor's nature and possibly suggests the existence of more classes.
The recent discovery that one of the repeaters, FRB\,180916.J0158+65, is periodic every $\sim16$~days with a short duty cycle, suggests a compact object, such as a neutron star (NS) belonging to a high-eccentricity binary system \citep{CHIME20}. However, the question remains as to what extent this source is representative of the observed sample.

Numerous theoretical models for the progenitors and for the radiation mechanism(s) have been proposed in the literature (see \citealt{Platts19} for a comprehensive review). The extreme brightness temperature ($T_b\ga 10^{35}$~K; \citealt{Lorimer07,Petroff19_rev}) combined with the ms duration and observed polarisation properties naturally suggest a coherent emission process from a compact source or from a relativistic expanding plasma. As large sources of rotational energy and strong magnetic fields, rapidly rotating NSs, or magnetars, either isolated or in binary systems, are among the most popular progenitor candidates.
Some of the radiation mechanisms proposed are (i) curvature emission by coherent bunches of charges \citep{Katz18}, that could result either from magnetic reconnection episodes close to the NS surface \citep{Kumar17,LuKumar18}, or from plasma instability triggered by clumpy ejecta within a binary black hole-massive star system \citep{Yi19}, or (ii) under specific conditions synchrotron maser emission \citep{Ghisellini17,LongPeer18,Metzger19,PlotnikovSironi19}.
While some of these radiation mechanisms predict no associated detectable prompt X/$\gamma$--ray emission \citep{GhiselliniLocatelli18}, the progenitor candidates are well known sources of high-energy flares.

In this context, FRBs could be powered by the huge magnetic fields of magnetars and could be associated with giant flares \citep{Popov10,Popov13,Beloborodov17}. Possible high-energy emission associated with FRBs could also be explained by the fact that magnetars are thought to form following either the core collapse of massive stars marked by long GRBs (L-GRBs; \citealt{Usov92,Thompson94,Bucciantini07,Metzger11}), or the merger of a binary neutron star (BNS) system marked by short GRBs (S-GRBs; \citealt{FanXu06,Metzger08}), or the accretion-induced collapse of a white dwarf \citep{Margalit19}.
The existence of repetitive FRB sources does not necessarily rule out cataclysmic models, in which the FRB is accompanied by the GRB itself either simultaneously or with some delay, due to the time it takes for a supramassive NS to finally collapse \citep{FalckeRezzolla14,Zhang14d}.

To date, FRB sources defied any search for associated hard X/soft $\gamma$--ray activity, in spite of an initial claim \citep{DeLaunay16}, which was not confirmed by a number of thorough, independent searches on different FRB samples \citep{Tendulkar16,Cunningham19,Martone19a} as well as on individual, exceptionally bright FRBs \citep{Guidorzi19}, and in some cases down to sub-second timescales \citep{Sun19,Anumarlapudi20}.
A search for prompt and afterglow emission in the $>$\,MeV energy range associated with two repetitive sources, one of which is the nearby FRB\,180916.J0158+65, ended up with no detection, either \citep{Casentini19}.

By reversing the strategy, \citet{Madison19} searched for FRBs from the directions of nearby short GRBs testing the possible existence of a young massive NS remnant capable of making FRBs, and found nothing down to the level of the faintest repetitions from FRB\,121102. \citet{Men19} carried out a similar analysis for six nearby (both long and short) GRBs with magnetar evidence and excluded a source with burst energy distribution and rate similar to 121102.

The Hard X-ray Modulation Telescope (HXMT), named ``Insight'' after launch on June 15, 2017, is the first Chinese X--ray astronomy satellite \citep{Li07,Zhang20_HXMT}. Its scientific payload consists of three main instruments: the  Low Energy X--ray telescope (LE; 1--15~keV; \citealt{Chen20_HXMT}), the Medium Energy X--ray telescope (ME; 5--30~keV; \citealt{Cao20_HXMT}), and the High Energy X--ray telescope (HE; \citealt{Liu20_HXMT}). The HE consists of 18 NaI/CsI detectors which cover the 20--250~keV energy band for pointing observations. In addition, it can be used as an open sky GRB monitor in the $0.2$--3~MeV energy range. The unique combination of a huge geometric area ($\sim5100$\,cm$^{2}$) and of continuous event tagging with timing accuracy $<10\,\mu$s, makes HXMT/HE an ideal instrument to search for possible $\gamma$--ray counterparts to FRBs down to ms or sub-ms scales in the keV--MeV energy range, where GRBs and magnetar giant flares release most of energy.
In this work we investigate this possibility by carrying out a systematic analysis of the data acquired with HE, used as an open sky $\gamma$--ray monitor.

Section~\ref{sec:dataset} describes the FRB sample; data analysis is reported in Section~\ref{sec:data_an}, whereas results are in Section~\ref{sec:res}. We discuss the implications in Section~\ref{sec:disc} and conclude in Section~\ref{sec:conc}.

Hereafter, we assume the latest Planck cosmological parameter
s: $H_0=67.74$~km\,s$^{-1}$\,Mpc$^{-1}$, $\Omega_m=0.31$, $\Omega_\Lambda=0.69$ \citep{cosmoPlanck15}.


\section{Data set}
\label{sec:dataset}
From the FRB catalogue \texttt{frbcat}\footnote{\url{http://www.frbcat.org}} \citep{Petroff16}, that contains nearly one hundred events (as of December 2019), we selected those which were visible from the {\em Insight}-HXMT location from the beginning of the mission (June 2017) to August 2019 and collected 43 FRBs. To this sample we added three recently discovered FRBs: 190711 \citep{UTMOST190711}, 190714 \citep{ASKAP190714}, and 190806 \citep{UTMOST190806}. This sample shrank from 46 to 39, since 7 occurred when the spacecraft was over the South Atlantic Anomaly (SAA) and no data are thus available. Hereafter, this will be referred to as the FRB sample. The selected FRBs were discovered by the following five telescopes: the Parkes radio telescope \citep{Oslowski19}, the Australian Square Kilometre Array Pathfinder (ASKAP; \citealt{Bannister17}), the upgraded Molonglo synthesis telescope (UTMOST; \citealt{Caleb17}),  the Canadian Hydrogen Intensity Mapping Experiment (CHIME; \citealt{CHIME19b}), and the Deep Synoptic Array ten-antenna prototype (DSA-10; \citealt{Ravi19b}).

In the present analysis we did not consider the as-yet most studied FRB, the repeater 121102, although it was visible for HXMT during the period of activity recorded on August 26, 2017 \citep{Gajjar18}. The reason is twofold: (i) the number of bursts is comparable with the sample itself and would strongly bias the results; (ii) there is independent evidence that it may not be representative of the observed population, based on the properties of its bursts (e.g., \citealt{James19}), and given also the different nature of its host galaxy with respect to the ones of the two one-off FRBs with known distance \citep{Tendulkar17,Bannister19,Ravi19b,Li19}. The recent discovery of a spiral galaxy 150\,Mpc away from Earth as the host of another repeater \citep{Marcote20}, along with the discovery of periodic patterns in the time history of its activity \citep{CHIME20}, adds to the case of the mysterious nature of FRB progenitors. Overall, as soon as HXMT/HE will observe FRB\,121102 and other repeaters during more periods of radio activity, a dedicated cumulative study for each of them is to be carried out.

For each FRB we checked the HXMT/HE operation mode and found that 9/39 (23\%) FRBs occurred during the GRB (low-gain) mode, a fraction which is somehow higher than that ($\sim 10$\,\%) of GRBs detected so far. With reference to the classification of repeating vs. one-off FRBs, in the light of recent results \citep{CHIME19c,Kumar19}, we also determined that 6/39 ($\sim 15$\%) are repeaters. Clearly, this number is likely to increase in the future, as soon as other FRBs, that are presently classified as one-off events, will be seen to repeat (e.g. \citealt{Ravi19c}).

For each FRB we calculated the local direction ($\theta$, $\phi$) with reference to the spacecraft frame. Given our interest in exploring the ms and sub-ms timescales, we had to determine the expected arrival time of each FRB at the spacecraft position. To this aim, from the FRB times in the FRB catalogue, that are taken from the literature and are usually reported at different frequencies, we calculated the corresponding arrival times referred at infinite frequency (that is, obtained after removing the delay due to the dispersion measure).\footnote{The delay due to DM was calculated as $\Delta\,t ({\rm ms})= 4.14702\, ({\rm DM}\,\nu^{-2})$, where DM is expressed in pc\,cm$^{-3}$ and $\nu$, expressed in GHz, is the reference frequency of the radio observation (eq.~1 of \citealt{Petroff19_rev}).} The calculated temporal shifts range from $0.4$ to $11.6$~s, with a mean (median) value of $2.5$~s ($1.4$~s). More importantly, the corresponding uncertainties, connected with the errors of the DM measures, are in the worst case as high as $178$~ms, with mean and median values of $3$ and $1.2$~ms, respectively.

Finally, we calculated the difference in the light travel time due to the relative position of the spacecraft with respect to the radio telescope that detected each FRB.\footnote{We made use of the python module {\tt astropy.time} (v.3.2.1) \citep{astropy:2013,astropy:2018}.}

Table~\ref{tab:alle} reports the details for all of the selected FRBs. In particular, both the UT of the detection by the radio telescope calculated at infinite frequency and the expected arrival UT at the spacecraft location are reported: they differ by a few ms, the largest delay being 29~ms, compatibly with expectations.\footnote{No wonder that the delay distribution is skewed toward negative delays (corresponding to an arrival time at the spacecraft preceding that at the radio telescope), since Earth-blocked FRBs (that is, when HXMT was on the other side of the Earth relative to the FRB direction) have already been rejected from the sample.}
Notably, the sample includes three FRBs with measured redshift: the periodic repeater 180916.J0158+65 ($z=0.0337$; \citealt{Marcote20,CHIME20}), and two one-off, 180924 ($z=0.3214$; \citealt{Bannister19}), and 190523 ($z=0.66$; \citealt{Ravi19b}).

\subsection{HXMT/HE data}
\label{sec:HEdata}
For each FRB, we extracted the event files, along with auxiliary files that include time-resolved information about the detectors' dead time, spacecraft's attitude and position, within a time window typically spanning from $-100$ to $300$~s around the FRB time. Since we used the HE units as an open-sky monitor, for each of the 18 HE detectors we extracted light curves (i.e., counts as a function of time) by selecting only the CsI events based on the pulse width. The light curves have both raw and background-subtracted counts within the total energy passband, which depends on the HE operation mode:
\begin{itemize}
    \item normal mode: 40--600~keV;
    \item GRB mode: 200-3000~keV.
\end{itemize}
The background was estimated through interpolation with a polynomial of up to third degree within two time windows, respectively preceding and following the interval that contains the FRB time. The size of each time window varies for different FRBs and had to be determined manually until a satisfactory subtraction was obtained.\footnote{We used the runs test to ensure the absence of trends in the background-subtracted rates.}

The rates in the light curves have been corrected for dead time effects. In practice, given the absence of intense peaks as the ones that are typically observed in the time profile of a bright GRB, dead time corrections never exceed 1--2\% and, as such, have a negligible impact. For this reason, when we considered integration times as short as 1~ms or even shorter, with a very few counts per bin, we did not apply any dead time correction and worked directly on the observed counts under the assumption of a Poisson distribution.

\begin{table*}
\centering
\caption{List of 39 FRBs that were promptly visible from {\em Insight}-HXMT and that were considered for the analysis of the present work.}
\label{tab:alle}
\begin{tabular}{lcccrcccrrr}
\hline
FRB & UT$^{\rm (a)}$ & R.A. & Decl. & Elev.$^{\rm (b)}$ & R$^{\rm (c)}$ & T$^{\rm (d)}$ & UT$_{\rm hxmt}^{\rm (e)}$ & $\theta^{\rm (f)}$ & $\phi^{\rm (g)}$ & M$^{\rm (h)}$\\
 &  & (J2000) & (J2000) & ($^\circ$) & & & & ($^\circ$) & ($^\circ$) &\\
\hline
170712               & 13:22:16.624 & 22:36:00.0 & $-60$:57:00 & $26.4$ & N & 1 & 13:22:16.617	 &  $120.1$ & $137.5$ & G\\
170906               & 13:06:55.527 & 21:59:48.0 & $-19$:57:00 & $13.9$ & N & 1 & 13:06:55.505	 &  $ 76.6$ & $ 95.3$ & G\\
171003               & 04:07:22.640 & 12:29:30.0 & $-14$:07:00 & $14.1$ & N & 1 & 04:07:22.616	 &  $113.0$ & $256.7$ & N\\
171004               & 03:23:38.501 & 11:57:36.0 & $-11$:54:00 & $66.3$ & N & 1 & 03:23:38.496	 &  $ 98.7$ & $260.1$ & N\\
171019               & 13:26:38.962 & 22:17:30.0 & $-08$:40:00 & $61.1$ & Y$^{\rm (i)}$ & 1 & 13:26:38.957	 &  $132.7$ & $118.0$ & N\\
171116               & 14:59:31.782 & 03:31:00.0 & $-17$:14:00 & $35.4$ & N & 1 & 14:59:31.767	 &  $112.1$ & $ 63.3$ & G\\
171209               & 20:34:20.298 & 15:50:25.0 & $-46$:10:20 & $82.1$ & N & 0 & 20:34:20.304	 &  $107.4$ & $274.4$ & N\\
171213               & 14:22:40.076 & 03:39:00.0 & $-10$:56:00 & $60.7$ & N & 1 & 14:22:40.070	 &  $ 52.4$ & $151.6$ & N\\
171216               & 17:59:10.322 & 03:28:00.0 & $-57$:04:00 & $37.9$ & N & 1 & 17:59:10.315	 &  $147.9$ & $ 14.1$ & N\\
180110               & 07:34:33.196 & 21:53:00.0 & $-35$:27:00 & $15.3$ & N & 1 & 07:34:33.172	 &  $144.9$ & $199.0$ & N\\
180119               & 12:24:29.756 & 03:29:18.0 & $-12$:44:00 & $89.8$ & N & 1 & 12:24:29.757	 &  $ 96.7$ & $166.0$ & N\\
180128.0             & 00:59:37.531 & 13:56:00.0 & $-06$:43:00 & $48.6$ & N & 1 & 00:59:37.528	 &  $ 28.1$ & $ 78.0$ & N\\
180128.2             & 04:53:25.575 & 22:22:00.0 & $-60$:15:00 & $62.9$ & N & 1 & 04:53:25.573	 &  $ 77.9$ & $223.0$ & N\\
180131               & 05:45:03.701 & 21:49:54.0 & $-40$:41:00 & $99.2$ & N & 1 & 05:45:03.703	 &  $101.4$ & $287.9$ & N\\
180301               & 07:34:18.627 & 06:12:43.4 & $+04$:33:45 & $19.0$ & N & 0 & 07:34:18.611	 &  $138.4$ & $155.1$ & N\\
180311               & 04:11:51.405 & 21:31:33.4 & $-57$:44:26 & $10.2$ & N & 0 & 04:11:51.385	 &  $ 44.1$ & $232.7$ & N\\
180430               & 09:59:58.049 & 06:51:00.0 & $-09$:57:00 & $43.5$ & N & 1 & 09:59:58.038	 &  $ 91.1$ & $207.2$ & N\\
180515               & 21:57:25.610 & 23:13:12.0 & $-42$:14:46 & $28.0$ & N & 1 & 21:57:25.593	 &  $ 61.9$ & $213.6$ & G\\
180525               & 15:19:05.559 & 14:40:00.0 & $-02$:12:00 & $97.1$ & N & 1 & 15:19:05.562	 &  $ 82.8$ & $ 61.4$ & N\\
180528               & 04:23:55.738 & 06:38:49.8 & $-49$:53:59 & $21.7$ & N & 2 & 04:23:55.717	 &  $113.5$ & $ 29.5$ & G\\
180714               & 10:00:05.481 & 17:46:12.0 & $-11$:45:47 & $52.5$ & N & 0 & 10:00:05.476	 &  $ 70.6$ & $109.9$ & N\\
180729.J0558+56      & 17:28:14.602 & 05:58:00.0 & $+56$:30:00 & $ 2.3$ & N & 3 & 17:28:14.573	 &  $113.8$ & $218.3$ & G\\
180730.J0353+87      & 03:37:16.156 & 03:53:00.0 & $+87$:12:00 & $56.6$ & N & 3 & 03:37:16.153	 &  $ 80.8$ & $233.5$ & N\\
180810.J0646+34      & 17:28:49.834 & 06:46:00.0 & $+34$:52:00 & $33.9$ & N & 3 & 17:28:49.818	 &  $155.6$ & $226.5$ & N\\
180810.J1159+83      & 22:40:40.545 & 11:59:00.0 & $+83$:07:00 & $29.1$ & N & 3 & 22:40:40.530	 &  $106.0$ & $194.4$ & N\\
180817.J1533+42      & 01:49:08.603 & 15:33:00.0 & $+42$:12:00 & $14.3$ & N & 3 & 01:49:08.578	 &  $ 17.8$ & $262.8$ & N\\
180916.J0158+65      & 10:15:15.779 & 01:58:00.0 & $+65$:44:00 & $58.1$ & Y & 3 & 10:15:15.772	 &  $132.5$ & $152.9$ & G\\
180923               & 04:03:34.037 & 15:10:55.4 & $-14$:06:10 & $ 1.6$ & N & 0 & 04:03:34.009	 &  $ 56.7$ & $291.7$ & G\\
180924               & 16:23:11.497 & 21:44:25.3 & $-40$:54:00 & $41.0$ & N & 1 & 16:23:11.487	 &  $ 52.5$ & $ 47.6$ & N\\
181017               & 10:24:36.023 & 22:05:54.8 & $-08$:50:34 & $83.0$ & N & 2 & 10:24:36.024	 &  $ 54.2$ & $ 32.7$ & N\\
181030.J1054+73      & 04:13:11.833 & 10:54:00.0 & $+73$:44:00 & $38.5$ & Y & 3 & 04:13:11.828	 &  $ 66.1$ & $341.1$ & N\\
181119.J12+65        & 16:48:58.996 & 12:42:00.0 & $+65$:08:00 & $12.3$ & Y & 3 & 16:48:58.971	 &  $ 63.4$ & $179.5$ & N\\
181228               & 13:48:48.067 & 06:09:23.6 & $-45$:58:02 & $31.8$ & N & 2 & 13:48:48.050	 &  $144.8$ & $ 53.4$ & G\\
190116.J1249+27      & 13:07:28.718 & 12:49:00.0 & $+27$:09:00 & $54.2$ & Y & 3 & 13:07:28.710	 &  $ 88.4$ & $155.8$ & N\\
190209.J0937+77      & 08:20:16.086 & 09:37:00.0 & $+77$:40:00 & $36.9$ & Y & 3 & 08:20:16.073	 &  $146.0$ & $ 48.8$ & N\\
190523               & 06:05:54.468 & 13:48:15.6 & $+72$:28:11 & $62.7$ & N & 4 & 06:05:54.465	 &  $ 83.6$ & $348.0$ & N\\
190711               & 01:53:39.575 & 21:56:00.0 & $-80$:23:00 & $70.0$ & N & 2 & 01:53:39.583	 &  $143.3$ & $ 76.5$ & N\\
190714               & 05:37:11.610 & 12:15:54.0 & $-13$:00:00 & $49.6$ & N & 1 & 05:37:11.607	 &  $133.1$ & $340.1$ & N\\
190806               & 17:07:55.689 & 00:02:21.4 & $-07$:34:55 & $71.3$ & N & 2 & 17:07:55.687	 &  $103.8$ & $ 56.9$ & N\\
\hline
\end{tabular}
\begin{list}{}{}
\item[$^{\rm (a)}$]{Detection UT at the radiotelescope site, referred to infinite frequency (that is, after removing the delay due to the dispersion measure).}
\item[$^{\rm (b)}$]{Elevation over the Earth limb of the FRB direction as observed from the spacecraft location.}
\item[$^{\rm (c)}$]{Repeating FRB: Yes/No (as up to the time of writing).}
\item[$^{\rm (d)}$]{Radiotelescope ID: 0=Parkes, 1=ASKAP, 2=UTMOST, 3=CHIME, 4=DSA-10.}
\item[$^{\rm (e)}$]{Expected arrival UT at the spacecraft location.}
\item[$^{\rm (f)}$]{Polar angle of the FRB direction with respect to the spacecraft frame.}
\item[$^{\rm (g)}$]{Azimuthal angle of the FRB direction with respect to the spacecraft frame.}
\item[$^{\rm (h)}$]{Operation mode of HXMT/HE: G (GRB), N (normal).}
\item[$^{\rm (i)}$]{Faint repetitions of this source have recently been reported \citep{Kumar19}.}
\end{list}
\end{table*}

\section{Data analysis}
\label{sec:data_an}
Searching for a transient signal over a large range of durations, from sub-ms to several seconds long, with no a-priori knowledge on the temporal structure, is optimally carried out by combining different, possibly complementary strategies. We therefore adopted three different approaches:
\begin{enumerate}
    \item we searched for $n\ge3$ simultaneous peaks in the counts of individual bins of as many HE detectors, significantly in excess of some given thresholds determined assuming Poisson distributions, whose expected values are given by the locally estimated background. Hereafter, this is referred to as the ``multi-detector search'';
    \item similarly to the previous case, the search for significant peaks was carried out on the total light curve, resulted from summing the counts of all 18 HE detectors. Hereafter, this is referred to as the ``summed-detector search'';
    \item we applied the peak search algorithm {\sc mepsa} \citep{Guidorzi15a} to 64-ms background-subtracted light curves. Originally conceived to identify peaks in GRB light curves over a large variety of durations and temporal structures, it was proved to perform better than other analogous algorithms. Hereafter, this is referred to as the ``{\sc mepsa} search''.
\end{enumerate}
For the first two methods we considered the following sequence of integration times: $100\,\mu$s, $1$\,ms, $10$\,ms, $64$\,ms, and $1.024$~s. In each case, for the first method the screened time window was $[-10,10]$~s, for which the background interpolation was sufficiently reliable and the number of bins to be screened and the consequent number of expected false positives remained manageable. Another reason that led us to exclude wider temporal windows for the search of coincident signals is the lack of information on the arrival direction of any possible transient candidate: only a strict temporal coincidence can reduce the chance probability of a fortuitously simultaneous unrelated transient, such as a GRB, and therefore remains essential to establish a possible association. Nonetheless, long duration (up to several ten seconds) candidates were screened through the {\sc mepsa} search, which was extended to the whole light curve.
To better explore the possibility of a precursor or delayed activity, for the summed-detector search we adopted a wider temporal window, $[-50, +80]$~s for all of the explored timescales, except for $100\,\mu$s for which we kept the shorter window adopted for the multi-detector search for the same reasons of manageability explained above. 

\subsection{Multi-detector search}
\label{sec:MDS}
In the multi-detector search we exploit the fact that HE consists of 18 independent detecting units, so that the simultaneous occurrence in different units naturally rejects both high-energy particle spikes and statistical flukes. The choice of at least three triggered detectors was the result of a tuning to limit the number of false positives. The values for the thresholds on the counts of a single bin of a single detector were determined in such a way so as to give a low number ($\la1$) of expected false positives for each FRB, taking into account the multiple trials connected with the number of bins to be analysed. In more detail, let $c_i$, $b_i$ the counts and expected background counts for the $i$-th bin. In the absence of signal, the probability for $c_i$ of exceeding a given threshold $k$ is given by the cumulative probability function $F^{\rm (pois)}_\lambda(n)$ of Poisson distribution,\footnote{This is the probability for a Poisson variate with $\lambda$ parameter of being $\le n$:  $F^{\rm (pois)}_\lambda(n) = \sum_{k=0}^n\ \lambda^k\,{\rm e}^{-\lambda}/k!$.} $p_{\rm sing}=1-F^{\rm (pois)}_{b_i}(k)$. The combined probability of the same event occurring simultaneously in at least $n$ out of 18 independent detectors is given by the surviving function of a binomial distribution, $p_{\rm comb}=1-F_{\rm binom}(n-1, 18, p_{\rm sing})$. The expected number of false positives is thus $N\,p_{\rm comb}$, where $N$ is the number of bins to be screened. The various threshold values used for $k$ are determined so as to have a given $p_{\rm sing}$, thus a given $p_{\rm comb}$. In practice, when dealing with very short integration times, that is in the very low count regime (where the expected counts can also be $<1$), due to the granularity of Poisson as a discrete distribution, both $p_{\rm sing}$ and $p_{\rm comb}$ can vary for each bin and the final number of expected false positives is the result of an average.

For each of the five explored integration times, we came up with the following threshold values on $p_{\rm sing}$, expressed in Gaussian $\sigma$ units:\footnote{This is just a common way of expressing the corresponding probabilities, although the interested distribution has nothing to do with a Gaussian. In other words, a threshold of $n\,\sigma$ is {\em not} equal to $n$ multiplied by the standard deviation of the corresponding Poisson distribution.} $2.8$ ($100\,\mu$s), $2.7$ (1\,ms), $2.6$ (10\,ms), $2.5$ (64\,ms), and $2.3$\,($1.024$~s). The decreasing threshold as a function of the increasing integration time reflects the decreasing number of bins.

\subsection{Summed-detector search}
\label{sec:SDS}
The summed-detector search is complementary to the former method especially for weak events, whose signal is not strong enough to trigger $n\ge3$ detectors simultaneously, but whose counts, summed over all the 18 detectors, is significant enough. Likewise, the threshold on the counts recorded in a single bin was chosen as to give a comparable probability to the combined one of the multi-detector search corresponding to the same temporal bin. With reference to the notation of Section~\ref{sec:MDS}, the threshold $k$ was chosen so that $p_{\rm sing} = 1 - F_{b_i}^{\rm (pois)}(k) \simeq p_{\rm comb}({\rm multi-detector})$, where $b_i$ is the background for the $i$-th bin of the total light curve. As a consequence, the summed-detector search provides a comparable number of false positives to that of the multi-detector one. Based on this, we set the corresponding threshold values on the single bin of the total light curve of any FRB: $4.8$ ($100\,\mu$s), $4.6$ (1\,ms), $4.1$ (10\,ms), $3.8$ (64\,ms), and $3.5$\,($1.024$~s).

We also applied this method to the cumulative light curve (that is, the sum of light curves of different FRBs) for the whole sample as well as for a number of sub-classes, such as repeating vs. non-repeating, or depending on the operation mode (and thus, energy passband) of the HE for each FRB.
Since this second application of the summed-detector search concerns just one (cumulative) light curve rather than 39 total light curves (one for each FRB), we expect a proportionally smaller number of false positives.

\subsection{{\sc mepsa} search}
\label{sec:mepsa}
This algorithm can only be applied to Gaussian noise background-subtracted time profiles. Because of this, we considered its use only for the total light curves of each individual FRB with an integration time of 64\,ms, which is long enough to ensure the Gaussian noise limit. Its advantage mainly relies in its capability of simultaneously exploring very different timescales and identifying the characteristic one of a given peak. Thanks to its versatility and to the relatively small number of bins to be screened compared to the other two methods, we applied it to the full light curve of each FRB, thus searching for several tens or hundred seconds long transients.

\section{Results}
\label{sec:res}
We did not find any candidates credibly associated with FRBs. Table~\ref{tab:summary} summarises the number of candidates for both multi- and summed-detector searches for the different integration times along with the corresponding number of expected false positives. In all cases the number of candidates is compatible with what is expected from the corresponding Poisson distribution.
Only for a couple of cases the number of candidates obtained from screening the total light curves is somehow higher than expected: for $\Delta\,t=100\,\mu$s, $\ge59$ candidates vs. an expected value of $45$ has a probability of $2.6$\%; for $\Delta\,t=10$\,ms, $\ge35$ candidates vs. $26$ expected has a probability of $5.3$\%. Visual inspection of the light curve of these candidates reveals no compelling evidence against the possibility that they are statistical flukes, though.
\begin{table*}
\centering
\caption{Summary of the results of the systematic search for $\gamma$--ray counterparts performed over a set of five different integration times through two different search methods: the multi-detector and the summed-detector searches (Sect.~\ref{sec:data_an}), applied to both the whole sample of FRBs and some of its sub-classes. For each pair of class/method and integration time, the number of candidates along with the expected number of false positives among parentheses is reported.}
\label{tab:summary}
\begin{tabular}{lccccc}
\hline
Class/search method & $\Delta\,t=100\,\mu$s & $\Delta\,t=1$\,ms & $\Delta\,t=10$\,ms & $\Delta\,t=64$\,ms & $\Delta\,t=1.024$\,s\\ 
\hline
Individual FRB: & & & & & \\
\ \ \ -\ multi-detector search$^{\rm (a)}$  &  20 ($23$) &  17 ($27$) &   8 ($14$) &  14 ($9$) & 7 ($4$)\\
\ \ \ -\ summed-detector search (SDS) $^{\rm (b)}$  & 59 ($45$) & 44 ($38$) & 35 ($26$) & 13 ($12$) &  3 ($2.3$)\\\hline
Cumulative LC: all (SDS)$^{\rm (c)}$ &   1 ($0.4$)  & 1 ($0.6$) & 0 ($0.56$) &     0 ($0.30$) &  0 ($0.06$)\\\hline
Cumulative LC: Rep/Non Rep (SDS)$^{\rm (d)}$: & & & & & \\
\ \ \ -\  Non Repetitive (33) &  1 ($0.4$) & 1 ($0.6$) & 0 ($0.56$) &  0 ($0.30$) &   0 ($0.06$)\\
\ \ \ -\  Repetitive      (6) &  1 ($0.9$) & 0 ($0.7$) & 0 ($0.58$) & 0 ($0.30$) & 1 ($0.06$)\\\hline
Cumulative LC: operation mode (SDS)$^{\rm (d)}$: & & & & & \\
\ \ \ -\  Normal  (30) & 0 ($0.4$) & 2 ($0.6$) & 1 ($0.56$) & 0 ($0.30$) & 0 ($0.06$)\\
\ \ \ -\  GRB   (9) & 3 ($0.66$) & 1 ($0.7$) & 2 ($0.58$)  & 0 ($0.30$) & 1 ($0.06$)\\
\hline
\end{tabular}
\begin{list}{}{}
\item[$^{\rm (a)}$]{Trigger condition: simultaneous trigger for $n\ge3$ out of the 18 HE detectors for any given FRB.}
\item[$^{\rm (b)}$]{Light curve obtained by summing the individual light curves of all 18 detectors for any given FRB.}
\item[$^{\rm (c)}$]{Light curve obtained by summing all the total light curves of all 39 FRBs.}
\item[$^{\rm (d)}$]{Number of FRBs per class is reported among parentheses.}
\end{list}
\end{table*}

Interestingly, for a couple of candidates found in the multi-detector search, one for FRB\,171209 at $t=-0.4$\,s ($\Delta\,t=64$\,ms) and another one for FRB\,180729.J0558+56 at $t=7.4$\,s ($\Delta\,t=1.024$\,s), four detectors simultaneously triggered the search. In the former case, the null hypothesis probability for such a single coincidence is $6.3\times10^{-5}$, which, multiplied by $3312$, the number of bins screened in a single $64$-ms light curve, and by $39$, the total number of FRBs, yields $0.77$ expected candidates. Similarly for the latter candidate, the corresponding null hypothesis probability is $5.1\times10^{-4}$, which, multiplied by 19 bins and by 39 FRBs, yields $0.38$ expected false positives. So both candidates are compatible with being rare simultaneous statistical flukes.

The {\sc mepsa} search has come up with a list of 6 candidates, whose details are reported in Table~\ref{tab:alleMEPSA}. All have significance between $4\sigma$ and $5\sigma$ (Gaussian), except for one ($5.2\sigma$) having the longest timescale ($\sim10$\,s), probably caused by the non-optimal background modelling as a consequence of unusually pronounced variability.  In none of these cases the FRB time is included in the time interval centred on the candidate with a duration comparable with the timescale identified by {\sc mepsa}. That half of them follow the FRB and the remaining precede it, suggests no systematic behaviour and is therefore compatible with the lack of any physical connection with FRBs.

\begin{figure*}
\centering
\includegraphics[width=9cm]{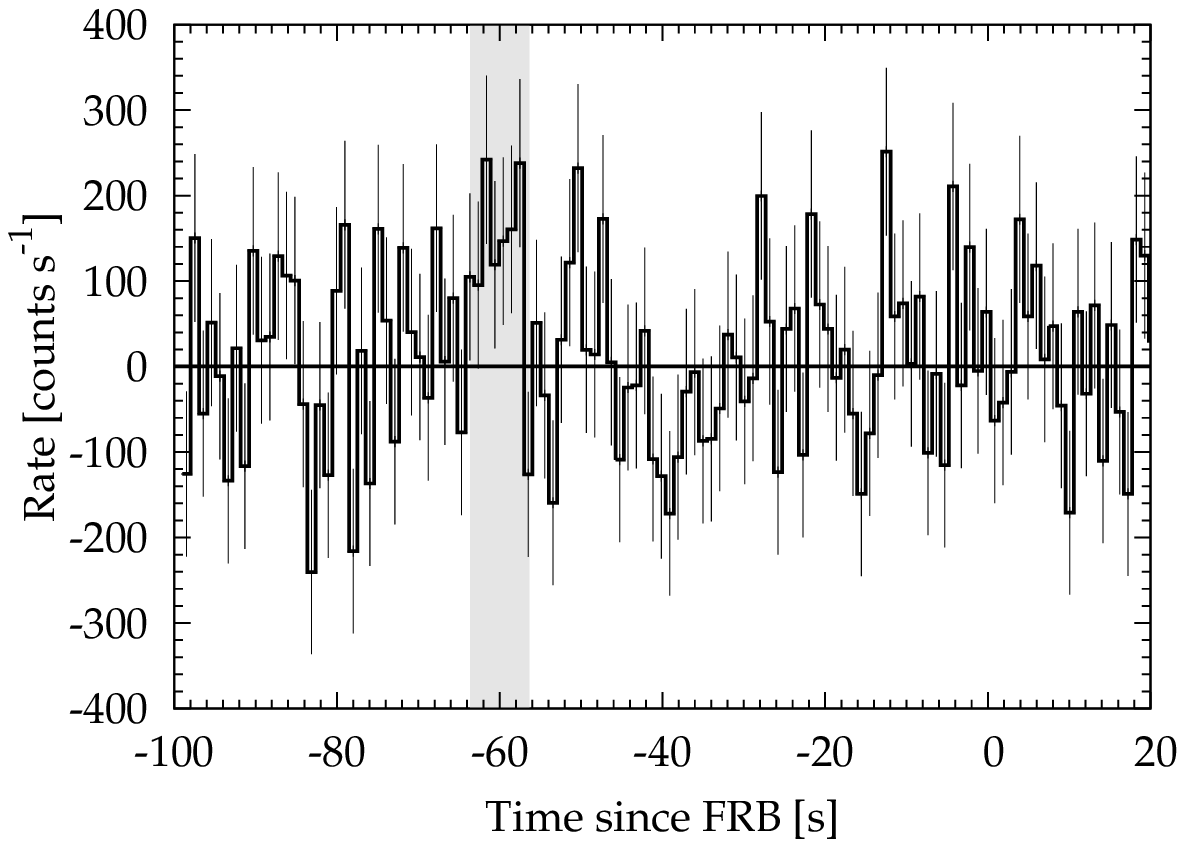}
\includegraphics[width=9cm]{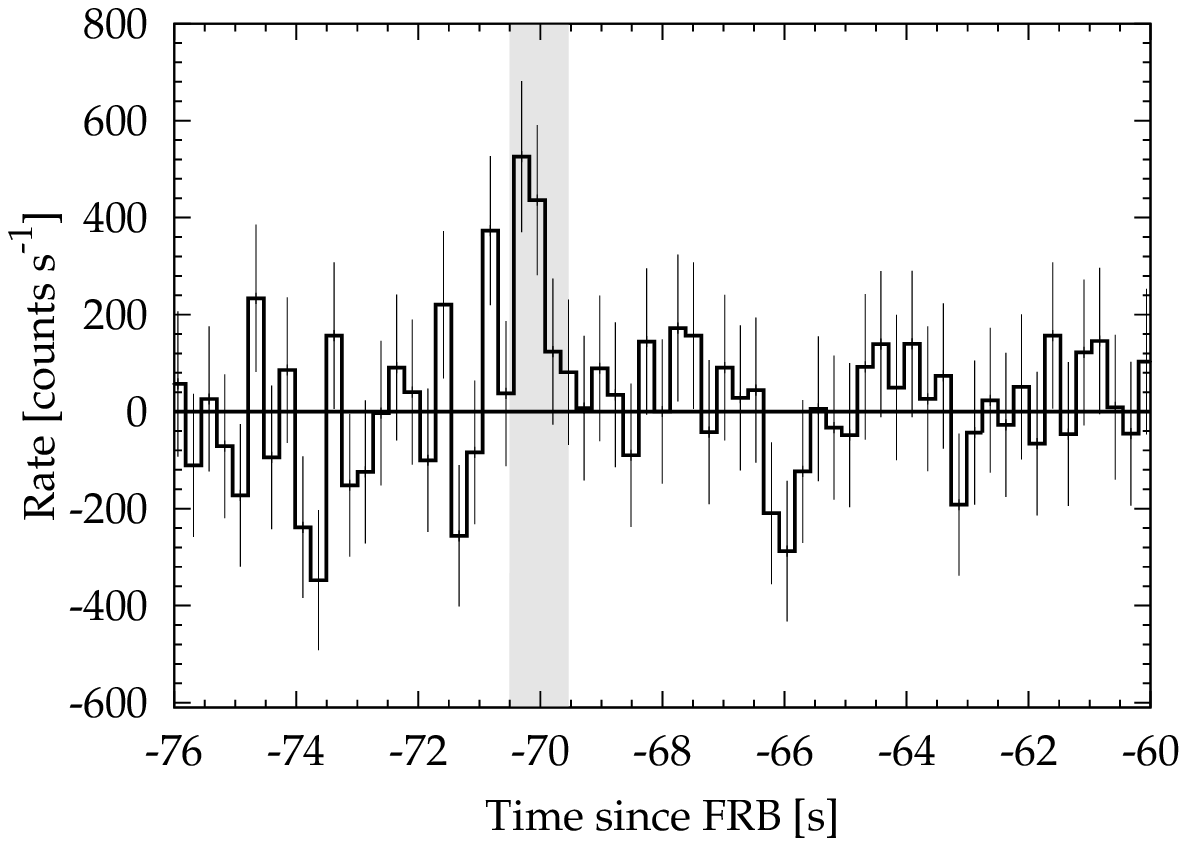}
\caption{Examples of background-subtracted light curves of two transient candidates found with {\sc mepsa}: one in 180525 ({\em left}) and the other in 180528 ({\em right}). Vertical shaded areas show the peak candidates identified through the {\sc mepsa} search. The two bin times are $1.024$ and $0.256$~s, respectively.}
\label{fig:mepsa_candidates}
\end{figure*}

Although it is unlikely that all of them are statistical fluctuations, the physical association with the FRB appears to be unsubstantiated. As an example, Figure~\ref{fig:mepsa_candidates} shows the time window of the total light curve for two FRBs each of which contains one transient candidate: FRB\,180525 and FRB\,180528. The analysis of the normalised background-subtracted counts, that is divided by the corresponding uncertainties, revealed no deviation from a standard normal distribution. This rules out the possibility that these candidates are the result of some unaccounted background variability.

\begin{table}
\caption{Candidates found with peak-search algorithm {\sc mepsa} in the 64-ms background-subtracted light curves summed over all 18 HE detectors.}
\label{tab:alleMEPSA}
\begin{tabular}{lrrrr}
\hline
FRB & $t^{\rm (a)}$ & $\Delta\,t^{\rm (b)}$ & $r^{\rm (c)}$ & S/N$^{\rm (d)}$\\
 &  (s) & (s) & (cts/s) & ($\sigma$) \\
\hline
FRB170712            & $105.892$  & $ 0.064$ &  $1607 \pm 321$ & $5.0$\\
FRB180311            & $-48.960$  & $ 0.384$ &  $ 494 \pm 137$ & $3.6$\\
FRB180525            & $-59.672$  & $ 6.336$ &  $ 179 \pm  39$ & $4.5$\\
FRB180528            & $-70.020$  & $ 0.832$ &  $ 372 \pm  85$ & $4.4$\\
FRB180923            & $ 21.876$  & $10.496$ &  $ 143 \pm  27$ & $5.2$\\
FRB190714            & $276.446$  & $ 1.472$ &  $ 364 \pm  85$ & $4.3$\\
\hline
\end{tabular}
\begin{list}{}{}
\item[$^{\rm (a)}$]{Time since FRB expected arrival time referenced to infinite frequency (i.e., the time delay due to dispersion is removed).}
\item[$^{\rm (b)}$]{Timescale of the peak candidate as evaluated with {\sc mepsa}.}
\item[$^{\rm (c)}$]{Background-subtracted count rate in the total HE passband (which depends on the operation mode).}
\item[$^{\rm (d)}$]{Signal-to-noise ratio.}
\end{list}
\end{table}
%

%
%

Having established that in no case we found a credible counterpart, for each event and for each of the explored timescales we calculated the corresponding upper limit on the average $\gamma$--ray flux as follows. We assumed a power-law spectrum with photon index $\Gamma=2$, which is representative of a non-thermal emission typically associated to GRBs or young NSs such as Crab. For each FRB and its local direction within the spacecraft reference frame, we then built the response functions for each of the 18 HE detectors from the mass model (v1) of the payload based on on-ground calibrations \citep{Liu20_HXMT}.
For any given FRB, the total-count-rate-to-flux conversion has been determined by summing up  the count rates of all detectors from the corresponding fake spectra generated under the same model with a given flux, and finally applied to the total net (i.e., background-subtracted) count rates corresponding to the average thresholds adopted in the summed-detector search. The resulting upper limits along with the corresponding significance values are reported in Table~\ref{tab:alleUL}. Clearly, the corresponding energy passband depends on the HE operation mode for each FRB.

In addition to the five timescales considered in the multi-detector and summed-detector searches, we also estimated the upper limits on a 10-s timescale using the {\sc mepsa} search as follows: we considered the $5.2$-$\sigma$ confidence candidate, whose characteristic time is $\sim10$~s. We modelled its profile and created some synthetic realisations of it, which were then added to the time profiles of other FRB light curves. We made sure that {\sc mepsa} detected all of these synthetic peaks. We therefore assumed the counts and peak count rate of this candidate as $5\sigma$ upper limits and converted to flux and fluence for each FRB using the corresponding count rate-to-flux conversion factor. This is a conservative estimate, as {\sc mepsa} can confidently detect fainter peaks. The results are reported in the last column of Table~\ref{tab:alleUL}.

\begin{table*}
\caption{Upper limits on the flux of possible $\gamma$--ray counterparts as evaluated on a set of five different integration times.}
\label{tab:alleUL}
\begin{tabular}{lrcccccc}
\hline
FRB & Passband & $F(100\,\mu{\rm s})^{\rm (a)}$ & $F(1\,{\rm ms})^{\rm (a)}$ & $F(10\,{\rm ms})^{\rm (a)}$ & $F(64\,{\rm ms})^{\rm (a)}$ & $F(1.024\,{\rm s})^{\rm (a)}$ & $F(10\,{\rm s})^{\rm (b)}$\\
 &  (keV) & ($10^{-5}$
 &($10^{-5}$
 &($10^{-6}$
 &($10^{-7}$
 &($10^{-8}$
 &($10^{-7}$\\
 &       & erg\,cm$^{-2}$s$^{-1}$) &   erg\,cm$^{-2}$s$^{-1}$)
 & erg\,cm$^{-2}$s$^{-1}$)
 & erg\,cm$^{-2}$s$^{-1}$)
 & erg\,cm$^{-2}$s$^{-1}$)
 & erg\,cm$^{-2}$s$^{-1}$) \\
\hline
170712            & $200$--$3000$ & $4.7$ ($4.3$) & $1.1$ ($4.5$) & $2.6$ ($4.0$) & $8.9$ ($3.8$) & $ 17$ ($3.5$) & $1.22$\\
170906            & $200$--$3000$ & $ 16$ ($4.7$) & $3.4$ ($4.4$) & $8.3$ ($4.1$) & $ 28$ ($3.8$) & $ 55$ ($3.5$) & $3.66$\\
171003            & $ 40$--$ 600$ & $3.3$ ($4.8$) & $0.69$ ($4.5$) & $1.7$ ($4.1$) & $5.9$ ($3.8$) & $ 12$ ($3.5$) & $0.65$\\
171004            & $ 40$--$ 600$ & $5.4$ ($4.7$) & $1.2$ ($4.5$) & $2.9$ ($4.1$) & $ 10$ ($3.8$) & $ 20$ ($3.5$) & $1.09$\\
171019            & $ 40$--$ 600$ & $2.6$ ($4.6$) & $0.57$ ($4.4$) & $1.4$ ($4.1$) & $  5$ ($3.8$) & $ 10$ ($3.5$) & $0.53$\\
171116            & $200$--$3000$ & $6.7$ ($4.6$) & $1.4$ ($4.4$) & $3.4$ ($4.0$) & $ 12$ ($3.8$) & $ 22$ ($3.5$) & $1.50$\\
171209            & $ 40$--$ 600$ & $3.2$ ($4.4$) & $0.76$ ($4.6$) & $1.8$ ($4.0$) & $6.2$ ($3.8$) & $ 12$ ($3.5$) & $0.73$\\
171213            & $ 40$--$ 600$ & $7.7$ ($4.4$) & $1.9$ ($4.5$) & $4.6$ ($4.1$) & $ 16$ ($3.8$) & $ 31$ ($3.5$) & $1.58$\\
171216            & $ 40$--$ 600$ & $2.6$ ($4.4$) & $0.63$ ($4.5$) & $1.5$ ($4.0$) & $5.3$ ($3.8$) & $ 10$ ($3.5$) & $0.60$\\
180110            & $ 40$--$ 600$ & $2.2$ ($4.4$) & $0.51$ ($4.5$) & $1.2$ ($4.0$) & $4.4$ ($3.8$) & $8.2$ ($3.5$) & $0.50$\\
180119            & $ 40$--$ 600$ & $4.9$ ($4.3$) & $1.2$ ($4.5$) & $2.9$ ($4.0$) & $ 10$ ($3.8$) & $ 19$ ($3.5$) & $1.12$\\
180128.0          & $ 40$--$ 600$ & $6.8$ ($4.6$) & $1.5$ ($4.5$) & $3.7$ ($4.1$) & $ 13$ ($3.8$) & $ 25$ ($3.5$) & $1.36$\\
180128.2          & $ 40$--$ 600$ & $7.7$ ($4.5$) & $1.8$ ($4.5$) & $4.5$ ($4.1$) & $ 16$ ($3.8$) & $ 30$ ($3.5$) & $1.57$\\
180131            & $ 40$--$ 600$ & $4.6$ ($4.8$) & $0.99$ ($4.5$) & $2.4$ ($4.1$) & $8.4$ ($3.8$) & $ 16$ ($3.5$) & $0.92$\\
180301            & $ 40$--$ 600$ & $2.6$ ($4.7$) & $0.56$ ($4.5$) & $1.4$ ($4.1$) & $4.8$ ($3.8$) & $9.1$ ($3.5$) & $0.52$\\
180311            & $ 40$--$ 600$ & $  7$ ($4.5$) & $1.6$ ($4.5$) & $3.8$ ($4.0$) & $ 13$ ($3.8$) & $ 25$ ($3.5$) & $1.58$\\
180430            & $ 40$--$ 600$ & $8.3$ ($4.8$) & $1.9$ ($4.6$) & $4.6$ ($4.1$) & $ 16$ ($3.8$) & $ 31$ ($3.5$) & $1.37$\\
180515            & $200$--$3000$ & $ 16$ ($4.6$) & $3.7$ ($4.5$) & $9.1$ ($4.0$) & $ 33$ ($3.8$) & $ 61$ ($3.5$) & $3.33$\\
180525            & $ 40$--$ 600$ & $7.7$ ($4.5$) & $1.8$ ($4.5$) & $4.4$ ($4.1$) & $ 16$ ($3.8$) & $ 30$ ($3.5$) & $1.56$\\
180528            & $200$--$3000$ & $6.4$ ($4.7$) & $1.3$ ($4.5$) & $3.2$ ($4.0$) & $ 11$ ($3.8$) & $ 21$ ($3.5$) & $1.43$\\
180714            & $ 40$--$ 600$ & $9.2$ ($4.6$) & $2.1$ ($4.5$) & $5.3$ ($4.1$) & $ 19$ ($3.8$) & $ 36$ ($3.5$) & $1.68$\\
180729.J0558+56   & $200$--$3000$ & $7.5$ ($4.6$) & $1.7$ ($4.5$) & $4.2$ ($4.1$) & $ 15$ ($3.8$) & $ 28$ ($3.5$) & $1.36$\\
180730.J0353+87   & $ 40$--$ 600$ & $6.9$ ($4.4$) & $1.6$ ($4.5$) & $3.9$ ($4.0$) & $ 13$ ($3.8$) & $ 26$ ($3.5$) & $1.57$\\
180810.J0646+34   & $ 40$--$ 600$ & $3.4$ ($4.6$) & $0.78$ ($4.5$) & $1.9$ ($4.1$) & $6.6$ ($3.8$) & $ 12$ ($3.5$) & $0.69$\\
180810.J1159+83   & $ 40$--$ 600$ & $  3$ ($4.5$) & $0.66$ ($4.4$) & $1.6$ ($4.0$) & $5.6$ ($3.8$) & $ 11$ ($3.5$) & $0.68$\\
180817.J1533+42   & $ 40$--$ 600$ & $5.5$ ($4.5$) & $1.2$ ($4.5$) & $  3$ ($4.1$) & $ 10$ ($3.8$) & $ 20$ ($3.5$) & $1.24$\\
180916.J0158+65   & $200$--$3000$ & $4.7$ ($4.5$) & $1.1$ ($4.4$) & $2.6$ ($4.0$) & $8.9$ ($3.8$) & $ 17$ ($3.5$) & $1.07$\\
180923            & $200$--$3000$ & $ 17$ ($4.4$) & $3.7$ ($4.4$) & $  9$ ($4.0$) & $ 31$ ($3.8$) & $ 60$ ($3.5$) & $3.53$\\
180924            & $ 40$--$ 600$ & $8.2$ ($4.7$) & $1.8$ ($4.5$) & $4.4$ ($4.0$) & $ 15$ ($3.8$) & $ 29$ ($3.5$) & $1.65$\\
181017            & $ 40$--$ 600$ & $8.9$ ($4.7$) & $1.9$ ($4.4$) & $4.8$ ($4.1$) & $ 17$ ($3.8$) & $ 32$ ($3.5$) & $1.61$\\
181030.J1054+73   & $ 40$--$ 600$ & $ 10$ ($4.5$) & $2.5$ ($4.5$) & $6.3$ ($4.1$) & $ 22$ ($3.8$) & $ 42$ ($3.5$) & $1.58$\\
181119.J12+65     & $ 40$--$ 600$ & $  9$ ($4.8$) & $  2$ ($4.5$) & $4.8$ ($4.1$) & $ 17$ ($3.8$) & $ 32$ ($3.5$) & $1.62$\\
181228            & $200$--$3000$ & $5.9$ ($4.6$) & $1.3$ ($4.5$) & $3.1$ ($4.0$) & $ 11$ ($3.8$) & $ 20$ ($3.5$) & $1.33$\\
190116.J1249+27   & $ 40$--$ 600$ & $7.6$ ($4.8$) & $1.6$ ($4.4$) & $3.9$ ($4.1$) & $ 14$ ($3.8$) & $ 26$ ($3.5$) & $1.51$\\
190209.J0937+77   & $ 40$--$ 600$ & $4.3$ ($4.7$) & $0.99$ ($4.5$) & $2.5$ ($4.1$) & $8.9$ ($3.8$) & $ 17$ ($3.5$) & $0.66$\\
190523            & $ 40$--$ 600$ & $7.4$ ($4.4$) & $1.8$ ($4.6$) & $4.3$ ($4.1$) & $ 15$ ($3.8$) & $ 29$ ($3.5$) & $1.50$\\
190711            & $ 40$--$ 600$ & $2.7$ ($4.5$) & $0.64$ ($4.5$) & $1.6$ ($4.1$) & $5.5$ ($3.8$) & $ 11$ ($3.5$) & $0.55$\\
190714            & $ 40$--$ 600$ & $2.6$ ($4.7$) & $0.58$ ($4.5$) & $1.4$ ($4.1$) & $  5$ ($3.8$) & $ 10$ ($3.5$) & $0.48$\\
190806            & $ 40$--$ 600$ & $4.4$ ($4.7$) & $0.92$ ($4.4$) & $2.3$ ($4.1$) & $  8$ ($3.8$) & $ 15$ ($3.5$) & $0.87$\\
\hline
\end{tabular}
\begin{list}{}{}
\item[$^{\rm (a)}$]{The corresponding significance in Gaussian $\sigma$ units is reported among parentheses.}
\item[$^{\rm (b)}$]{Estimated through the {\sc mepsa} search. All are at $5\sigma$ confidence level.}
\end{list}
\end{table*}

\subsection{Upper limits on average luminosities}
\label{sec:upplim_z}
Except for the three FRBs with measured redshift $z$ that are included in the sample, for the remaining 36 we estimated upper limits on $z$ (95\% confidence) using their DM values, following the prescriptions by \citet{Pol19} and using their  code.\footnote{\url{https://github.com/NihanPol/DM_IGM}}
For each direction, these authors calculate the Galactic contribution to the observed DM using the NE2001 Galactic free electron density model \citep{Cordes02}, including the contribution of the Galactic halo, which is assumed in the range 50--80\,pc\,cm$^{-3}$ \citep{ProchaskaZheng19}. The DM contribution due to the intergalactic medium (IGM) is estimated by integrating the free electron density along the line of sight to the FRB derived from cosmological simulations of the evolution of large scale structures through dark matter particles in a redshift range $0 < z < 1.4$. Then, they converted the dark matter particle number density to baryonic matter density. Here we use the values obtained by weighting by the matter distribution (see \citealt{Pol19} for details).

For each upper limit on the average flux reported in Table~\ref{tab:alleUL} we calculated the corresponding upper limit on the average luminosity: the results are reported in Table~\ref{tab:alleULLum} and refer to the same energy passbands, integration times, and significance values of Table~\ref{tab:alleUL}. 

It is worth calculating the corresponding limits on the released energy over the ms and sub-ms timescales for the three FRBs with measured redshift: for 180916.J0158+65, $E(100\,\mu{\rm s})<1.5\times10^{46}$~erg, $E(1\,{\rm ms})<3.0\times10^{46}$~erg; for 180924, $E(100\,\mu{\rm s})<2.6\times10^{48}$~erg, $E(1\,{\rm ms})<5.2\times10^{48}$~erg; for 190523, 
$E(100\,\mu{\rm s})<1.0\times10^{49}$~erg, $E(1\,{\rm ms})<2.2\times10^{49}$~erg.

For each FRB we compared the luminosity of the radio pulse with the corresponding upper limit on the $\gamma$--ray counterpart evaluated on 1-ms timescale, assuming for both the redshift value given in Table~\ref{tab:alleULLum}. Consequently, the ratio between the two luminosities does not depend on $z$. The radio luminosity is calculated by multiplying the specific luminosity by the reference frequency of each radio telescope (that is, for the radio $L=\nu\,L_\nu$). The specific luminosity was calculated either directly from the flux density --when available-- or from the combination of fluence density and burst temporal width, as reported in \texttt{frbcat}. The resulting upper limits on the $L_{\gamma}/L_{\rm radio}$ ratios as well as the related quantities are reported in Table~\ref{tab:radio_ratio}. Clearly, for all FRBs we can exclude only cases in which the high-energy counterpart carries most energy by 6 to 10 orders of magnitude.

\begin{figure}
\centering
\includegraphics[width=9cm]{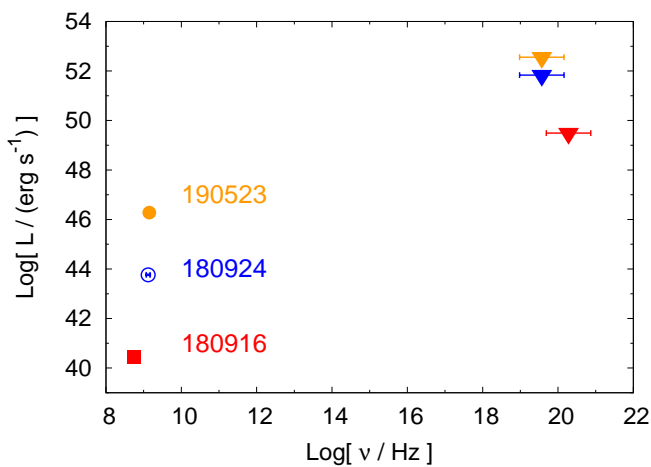}
\caption{FRB radio luminosity and 1-ms $\gamma$--ray upper limits for the three FRBs with measured redshifts.}
\label{fig:Lum_radiogamma_knownz}
\end{figure}

Given that for the three FRBs with measured $z$ the radio luminosity can be estimated, Figure~\ref{fig:Lum_radiogamma_knownz} displays their luminosities as a function of frequency as illustrative examples. More quantitatively, assuming a power-law dependence of the specific luminosity as a function of frequency, $L_\nu\propto\nu^{\alpha}$, we can constrain the mean value of the power-law index between radio and $\gamma$-rays, $\alpha+1 = \log{(L_\gamma/L_{\rm radio})}/\log{(\nu_\gamma/\nu_{\rm radio})}$.  The information encoded in the upper limits on the luminosity ratios (Table~\ref{tab:radio_ratio}) is the same as the one expressed in terms of $\alpha$, so the latter can be constrained for all FRBs. Here we simply report the upper limits on $\alpha$ for the three FRBs with known $z$: $\alpha<-0.21$ (180916.J0158+65), $\alpha<-0.23$ (180924), $\alpha<-0.40$ (190523).
%
%

%
\begin{table*}
\caption{Upper limits on the average luminosity of possible $\gamma$--ray counterparts as evaluated on a set of different integration times.}
\label{tab:alleULLum}
\begin{tabular}{llcccccc}
\hline
FRB & $z_{\rm max}^{\rm (a)}$ & $L(100\,\mu{\rm s})$ & $L(1\,{\rm ms})$ & $L(10\,{\rm ms})$ & $L(64\,{\rm ms})$ & $L(1.024\,{\rm s})$ & $L(10\,{\rm s})$\\
 &  & ($10^{53}$
 &($10^{52}$
 &($10^{51}$
 &($10^{51}$
 &($10^{50}$
 &($10^{50}$\\
 &       & erg\,s$^{-1}$) &   erg\,s$^{-1}$)
 & erg\,s$^{-1}$)
 & erg\,s$^{-1}$)
 & erg\,s$^{-1}$)
 & erg\,s$^{-1}$) \\
\hline
170712            & $0.6740$ & $0.99$ & $2.31$ & $5.45$ & $1.87$ & $4.30$ & $2.56$ \\
170906            & $0.7500$ & $4.35$ & $9.25$ & $22.60$ & $7.65$ & $17.50$ & $9.96$ \\
171003            & $0.8080$ & $1.08$ & $2.25$ & $5.55$ & $1.94$ & $4.47$ & $2.12$ \\
171004            & $0.6650$ & $1.10$ & $2.43$ & $5.88$ & $2.09$ & $4.76$ & $2.21$ \\
171019            & $0.8080$ & $0.85$ & $1.86$ & $4.57$ & $1.63$ & $3.83$ & $1.73$ \\
171116            & $0.9140$ & $2.96$ & $6.19$ & $15.00$ & $5.25$ & $11.70$ & $6.63$ \\
171209            & $1.2210$ & $2.89$ & $6.87$ & $16.30$ & $5.65$ & $12.30$ & $6.58$ \\
171213            & $0.4390$ & $0.58$ & $1.42$ & $3.44$ & $1.17$ & $2.70$ & $1.18$ \\
171216            & $0.5340$ & $0.31$ & $0.75$ & $1.79$ & $0.63$ & $1.40$ & $0.71$ \\
180110            & $0.9670$ & $1.12$ & $2.59$ & $6.10$ & $2.22$ & $4.91$ & $2.53$ \\
180119            & $0.7620$ & $1.39$ & $3.40$ & $8.21$ & $2.87$ & $6.36$ & $3.17$ \\
180128.0          & $0.7980$ & $2.15$ & $4.75$ & $11.70$ & $4.11$ & $9.28$ & $4.31$ \\
180128.2          & $0.8310$ & $2.69$ & $6.30$ & $15.70$ & $5.47$ & $12.30$ & $5.49$ \\
180131            & $0.9340$ & $2.15$ & $4.62$ & $11.20$ & $3.94$ & $8.65$ & $4.28$ \\
180301            & $0.7660$ & $0.75$ & $1.60$ & $4.01$ & $1.39$ & $3.08$ & $1.49$ \\
180311            & $1.2970$ & $7.34$ & $16.80$ & $39.80$ & $13.90$ & $30.70$ & $16.60$ \\
180430            & $0.3850$ & $0.46$ & $1.04$ & $2.52$ & $0.86$ & $1.98$ & $0.75$ \\
180515            & $0.7230$ & $3.98$ & $9.21$ & $22.60$ & $8.16$ & $18.00$ & $8.29$ \\
180525            & $0.7540$ & $2.12$ & $4.96$ & $12.10$ & $4.31$ & $9.69$ & $4.30$ \\
180528            & $1.0450$ & $3.94$ & $8.00$ & $19.70$ & $6.82$ & $15.00$ & $8.80$ \\
180714            & $1.2260$ & $8.39$ & $19.20$ & $48.40$ & $17.10$ & $38.30$ & $15.30$ \\
180729.J0558+56   & $0.6150$ & $1.26$ & $2.85$ & $7.05$ & $2.46$ & $5.57$ & $2.28$ \\
180730.J0353+87   & $1.0260$ & $4.06$ & $9.41$ & $22.90$ & $7.90$ & $17.80$ & $9.23$ \\
180810.J0646+34   & $0.7110$ & $0.81$ & $1.86$ & $4.54$ & $1.57$ & $3.50$ & $1.64$ \\
180810.J1159+83   & $0.4440$ & $0.23$ & $0.51$ & $1.23$ & $0.43$ & $0.97$ & $0.52$ \\
180817.J1533+42   & $1.1120$ & $3.94$ & $8.61$ & $21.50$ & $7.51$ & $16.80$ & $8.89$ \\
180916.J0158+65   & $0.0337^{\rm (b)}$ & $1.3\times10^{-3}$ & $3.1\times10^{-3}$ & $7.3\times10^{-3}$ & $2.5\times10^{-3}$ & $5.5\times10^{-3}$ & $3.0\times10^{-3}$\\
180923            & $0.8630$ & $6.53$ & $14.20$ & $34.60$ & $12.00$ & $27.00$ & $13.60$ \\
180924            & $0.3214^{\rm (c)}$ & $0.29$ & $0.65$ & $1.58$ & $0.55$ & $1.23$ & $0.59$ \\
181017            & $0.5870$ & $1.33$ & $2.85$ & $7.20$ & $2.58$ & $5.71$ & $2.41$ \\
181030.J1054+73   & $0.1000$ & $0.03$ & $0.07$ & $0.17$ & $0.06$ & $0.13$ & $0.04$ \\
181119.J12+65     & $0.7300$ & $2.29$ & $5.09$ & $12.20$ & $4.38$ & $9.70$ & $4.13$ \\
181228            & $0.6960$ & $1.34$ & $2.95$ & $7.03$ & $2.41$ & $5.31$ & $3.02$ \\
190116.J1249+27   & $0.8080$ & $2.48$ & $5.23$ & $12.70$ & $4.49$ & $9.89$ & $4.93$ \\
190209.J0937+77   & $0.7720$ & $1.26$ & $2.89$ & $7.30$ & $2.60$ & $5.71$ & $1.93$ \\
190523            & $0.66^{\rm (d)}$ & $1.47$ & $3.59$ & $8.57$ & $3.05$ & $6.81$ & $2.99$ \\
190711            & $0.8860$ & $1.11$ & $2.62$ & $6.55$ & $2.24$ & $5.20$ & $2.27$ \\
190714            & $0.8390$ & $0.93$ & $2.08$ & $5.02$ & $1.79$ & $4.20$ & $1.70$ \\
190806            & $0.7550$ & $1.22$ & $2.54$ & $6.36$ & $2.20$ & $4.86$ & $2.40$ \\
\hline
\end{tabular}
\begin{list}{}{}
\item[$^{\rm (a)}$]{Upper limit on redshift $z$ (95\% CL) calculated following the prescriptions of \citet{Pol19}.}
\item[$^{\rm (b)}$]{Spectroscopic redshift of the host galaxy \citep{Marcote20}.}
\item[$^{\rm (c)}$]{Spectroscopic redshift of the host galaxy \citep{Bannister19}.}
\item[$^{\rm (d)}$]{Spectroscopic redshift of the host galaxy \citep{Ravi19b}.}
\end{list}
\end{table*}

\begin{table*}
\caption{Upper limits on the $\gamma$-to-radio luminosity ratio, $L_{\gamma,{\rm 1\,ms}}/L_{\rm radio}$. Radio data (central frequency, flux density), radio luminosity calculated as $L_{\rm radio}=\nu\,L_\nu$ are also reported. Radio luminosities were calculated assuming the same redshift limits or values for the $\gamma$--ray ones, so their ratio does not depend on $z$.  For the FRBs for which the flux density is not reported in \texttt{frbcat}, we estimated it from the fluence density and the burst width.}
\label{tab:radio_ratio}
\begin{tabular}{lrrcr}
\hline
FRB & $\nu_{\rm radio}$ & $F_{\rm radio}$ & $L_{\rm radio}$ &  $\log{(L_{\gamma,{\rm 1 \,ms}}/L_{\rm radio})}$\\
 & (GHz)
 & (Jy)
 & (erg/s)
 & \\
\hline
170712  &          $1.297$ &   $37.8$ &  $1.0\times10^{45}$ &  $7.4$\\
170906  &          $1.297$ &   $29.6$ &  $1.0\times10^{45}$ &  $7.9$\\
171003  &          $1.297$ &   $40.5$ &  $1.7\times10^{45}$ &  $7.1$\\
171004  &          $1.297$ &   $22.0$ &  $5.8\times10^{44}$ &  $7.6$\\
171019  &          $1.297$ &   $40.5$ &  $1.7\times10^{45}$ &  $7.0$\\
171116  &          $1.297$ &   $19.6$ &  $1.1\times10^{45}$ &  $7.7$\\
171209  &          $1.352$ &   $1.48$ &  $1.8\times10^{44}$ &  $8.6$\\
171213  &          $1.297$ &   $88.6$ &  $8.6\times10^{44}$ &  $7.2$\\
171216  &          $1.297$ &   $21.0$ &   $3.3\times10^{44}$  &  $7.4$\\
180110  &          $1.297$ &   $128.1$ &   $8.4\times10^{45}$  &  $6.5$\\
180119  &          $1.297$ &   $40.7$ &   $1.5\times10^{45}$  &  $7.4$\\
180128.0&          $1.297$ &   $17.5$ &   $7.2\times10^{44}$  &  $7.8$\\
180128.2&          $1.297$ &   $28.7$ &   $1.3\times10^{45}$  &  $7.7$\\
180131  &          $1.297$ &   $22.2$ &   $1.3\times10^{45}$  &  $7.5$\\
180301  &          $1.352$ &   $1.30$ &   $5.0\times10^{43}$  & $8.5$\\
180311  &          $1.352$ &   $0.15$ &   $2.1\times10^{43}$  &  $9.9$\\
180430  &          $1.297$ &   $147.5$ &   $1.0\times10^{45}$  &  $7.0$\\
180515  &          $1.320$ &   $24.2$ &   $7.9\times10^{44}$   & $8.1$\\
180525  &          $1.297$ &   $78.9$ &   $2.8\times10^{45}$   & $7.2$\\
180528  &          $0.835$ &   $15.75$ &   $8.1\times10^{44}$   & $8.0$\\
180714  &          $1.352$ &   $0.6$ &   $7.4\times10^{43}$   & $9.4$\\
180729.J0558+56 &  $0.600$ &   $112.5$ &   $1.1\times10^{45}$  &  $7.4$\\
180730.J0353+87 &  $0.600$ &   $119.0$ &   $4.2\times10^{45}$  & $7.4$\\
180810.J0646+34 &  $0.600$ &   $40.74$ &   $5.8\times10^{44}$  & $7.5$\\
180810.J1159+83 &  $0.600$ &   $60.71$ &   $2.8\times10^{44}$   & $7.3$\\
180817.J1533+42 &  $0.600$ &   $70.27$ &   $3.0\times10^{45}$  &  $7.5$\\
180916.J0158+65 &  $0.600$ &   $1.64$ &   $2.8\times10^{40}$  &  $9.0$\\
180923      &      $1.352$ &   $2.90$ &   $1.5\times10^{44}$  &  $9.0$\\
180924      &      $1.320$ &   $12.3$ &   $5.8\times10^{43}$  &  $8.0$\\
181017      &      $0.835$ &   $161.$ &   $2.0\times10^{45}$  &  $7.2$\\
181030.J1054+73 &  $0.600$ &   $12.37$ &   $2.0\times10^{42}$  &  $8.5$\\
181119.J12+65   &  $0.600$ &   $0.29$ &   $4.4\times10^{42}$  & $10.1$\\
181228          &  $0.835$ &   $19.23$ &   $3.6\times10^{44}$  &  $7.9$\\
190116.J1249+27 &  $0.600$ &   $0.2$ &   $3.9\times10^{42}$  & $10.1$\\
190209.J0937+77 &  $0.600$ &   $0.54$ &   $9.5\times10^{42}$  &  $9.5$\\
190523          &  $1.411$ &   $667$ &   $1.9\times10^{46}$   & $6.3$\\
190711          &  $0.835$ &   $6.7^{\rm (a)}$ &   $2.3\times10^{44}$  & $8.1$\\
190714          &  $1.297$ &   $8.0^{\rm (a)}$ &   $3.7\times10^{44}$  & $7.7$\\
190806          &  $0.835$ &   $3.9$ &   $9.0\times10^{43}$  & $8.4$\\
\hline
\end{tabular}
\begin{list}{}{}
\item[$^{\rm (a)}$]{Only the fluence density is reported and no information is available on the burst width at the time of writing. A nominal value of 1\,ms was assumed.}
\end{list}
\end{table*}

\section{Discussion}
\label{sec:disc}
\begin{figure*}
\centering
\includegraphics[width=\textwidth]{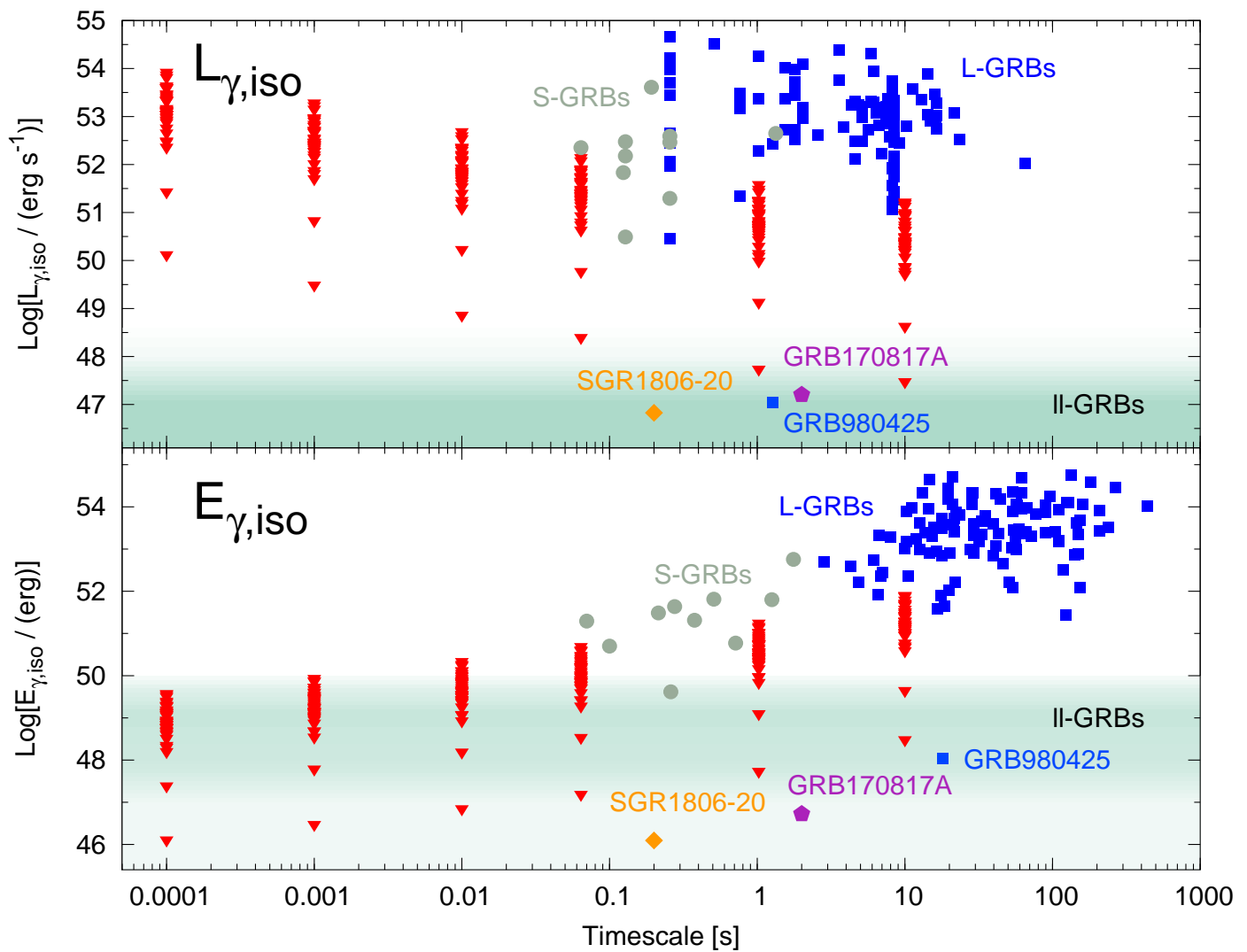}
\caption{{\em Top panel}: upper limits on average isotropic--equivalent luminosities of potential $\gamma$--ray counterparts to FRBs as a function of integration time (red upside down triangles). For comparison L-GRBs (blue squares) and S-GRBs (grey circles) are also shown (from \citealt{KWGRBcat17}), along with the subluminous short GRB\,170817A associated with GW\,170817 (purple pentagon) and the giant flare of Galactic magnetar SGR\,1806-20 (orange diamond). Low-luminosity GRBs populate the shaded area, with prototypical GRB\,980425 explicitly shown (light blue square). Uncertainties on luminosities have comparable sizes with symbols. {\em Bottom panel}: same as the top panel, with isotropic-equivalent radiated energy $E_{\gamma,{\rm iso}}$ instead of isotropic-equivalent luminosity. GRB timescales here correspond to $T_{90}$ durations.}
\label{fig:Lum_comparison}
\end{figure*}

Figure~\ref{fig:Lum_comparison} shows the upper limits on the (isotropic--equivalent) $\gamma$--ray luminosities and energies compared with the analogous values for the sample of L-GRBs and S-GRBs of the Konus-{\em Wind} catalogue \citep{KWGRBcat17}. The timescales used for the GRB luminosities refer to the integration time around the peak of each GRB, whereas for the energies it is just the duration measured with $T_{90}$. Our values clearly show that almost all of the observed populations of both L-GRBs and S-GRBs are incompatible with being simultaneously associated with FRBs, except for the subluminous cases.
%
%

By way of example, we show GRB\,980425 \citep{Pian00,Amati08,Yonetoku10}, the first long low-luminosity GRB ({\em ll}-GRB) discovered and still one of the nearest ones yet detected. {\em ll}-GRBs \citep{Kulkarni98,Campana06,Waxman07} have $E_{\gamma,{\rm iso}}\sim10^{48}$--$10^{50}$~erg, in the local universe they outnumber more energetic long GRBs  by a factor of $\sim100$ (ignoring probable differences in beaming factors), and their prompt emission likely has a different origin  \citep{Liang07,Amati07,Virgili09,WandermanPiran10,HowellCoward13}. Even the most stringent upper limit obtained from our sample cannot rule out a {\em ll}-GRB like GRB\,980425, although similar GRBs only a factor of few more luminous are excluded.

Figure~\ref{fig:Lum_comparison} also shows the subluminous short GRB\,170817A associated with the first BNS merger detected with gravitational interferometers, GW\,170817, which had $L_{\gamma,{\rm iso}}=(1.6\pm0.6)\times10^{46}$~erg\,s$^{-1}$ and $E_{\gamma,{\rm iso}}=(5.3\pm1.0)\times10^{46}$~erg \citep{LIGO-Fermi17,Goldstein17}. Should analogous subluminous S-GRBs have been associated with FRBs, our searches would not have detected them, as was also the case for HXMT/HE with GRB\,170817A, particularly because of its spectral softness \citep{Li17_GW170817}.

Cataclysmic models that associate FRBs with BNS mergers can be compatible with our results, only if X/$\gamma$--rays are much more collimated along the off-axis jet than radio emission \citep{Totani13}. Assuming that each of the one-off FRBs of our sample was associated to a BNS source like GW\,170817, whose structured jet had an opening angle of $4$--$6^{\circ}$ \citep{Hajela19,WuMacFadyen19,Troja19}, the beaming factor $f_b^{-1} = (1-\cos{\theta_j})^{-1}$ is in the range $200$--$400$, i.e. broadly compatible with our results. Although it is loosely constrained, the BNS volumetric rate, as estimated from the recent results of the first two runs of LIGO and Virgo interferometers \citep{LIGO-Virgo19_prx}, lies in the range $110$--$3840$~Gpc$^{-3}$\,yr$^{-1}$ (90\% confidence), that is roughly compatible with that expected for the non-repetitive fraction of the observed FRB population, whose total rate is $\ga10^4$~Gpc$^{-3}$\,yr$^{-1}$ (e.g., \citealt{Ravi19c}).

In the context of non-cataclysmic models, in Figure~\ref{fig:Lum_comparison} we compare our results also with the giant flare observed from Galactic magnetar SGR\,1806-20, whose initial $0.2$-s long spike had an isotropic-equivalent peak luminosity of $7\times10^{46}$\,erg\,s$^{-1}$ and energy of $(1.2\pm0.3)\times10^{46}$~erg \citep{Hurley05,Palmer05,Bibby08}.
While we cannot exclude the systematic occurrence of extra-galactic giant flares associated with our FRB sample, in the case of the nearest FRB\,180916.J0158+65, our limits are very close to it.

Another non-disruptive model describes the FRB mechanism in terms of  synchrotron maser emission at magnetised relativistic shocks, possibly caused by ejecta of young magnetars \citep{Metzger19}. In this model, such shocks are the result of relativistic shells emitted by the magnetar, impacting on a subrelativistic electron-ion outflow. In addition to generating the synchrotron maser emission, the same shock produces a down-scaled version of GRB afterglows, that peak in the hard X/$\gamma$--ray band over a timescale comparable with the FRB itself or somehow longer. Looking at Fig.~8 of \citet{Metzger19}, in the HXMT/HE energy band the predicted $\gamma$--ray luminosity lies in the range $10^{44}$--$10^{45}$~erg/s and lasts several ten ms. Our limits on such emission are still orders of magnitude above and only future generation detectors will be able to test this possibility.

\section{Conclusions}
\label{sec:conc}
Using the data of the open sky $\gamma$--ray monitor HE aboard HXMT, we constrained the prompt hard X/$\gamma$--ray emission in the keV--MeV band potentially associated with 39 FRBs over a range of timescales from $100$\,$\mu$s to several ten seconds. Using the measured redshifts for three FRBs and conservative upper limits on the redshifts of the remaining sources based on the observed DM, we derived upper limits on isotropic-equivalent $\gamma$--ray luminosities and released energies.

As long as one-off events still represent a sizeable fraction of the observed FRB population, cataclysmic models cannot be ruled out. In this context, we can confidently discard a systematic association of one-off FRBs with standard cosmological GRBs, both long and short. Conversely, subluminous GRBs instead cannot be rejected. Under the assumption that subluminous GRBs are standard energetic GRBs viewed off axis \citep{Ghisellini06,Salafia16}, and that FRBs are associated with them but less collimated, in the near future, when the FRB sample will be large enough to have at least a few on-axis cases, we should expect to see some of them with associated GRB emission. This prediction holds true for both core collapse of massive stars that are connected with L-GRBs, and for BNS mergers connected with S-GRBs. In the future, thanks to its broad energy band the mission concept Transient High-Energy Sky and Early Universe Surveyor (THESEUS; $0.3$--$10$~MeV; \citealt{Amati18_THESEUS}) will help clarify the nature of low-luminosity GRBs, most of which are spectrally soft.

Alternatively, if the radio and the possible high-energy emission have comparable beaming factors and timescales, our results still allow for a $L_{\gamma,{\rm iso}}<10^{49}$~erg/s on ms scales, corresponding to a process which releases more energy at high frequencies than in the radio by orders of magnitude.

Concerning non-cataclysmic FRB models, our results cannot reject giant flares from extra-galactic magnetars similar to those observed so far from Galactic siblings and only for the nearest FRB yet discovered, at $\sim150$~Mpc, we can rule out giant flares at least ten times more energetic than that of Galactic magnetar SGR\,1806-20.

\begin{acknowledgements}
We thank the anonymous referee for helping us improve the paper.
This work is supported by the National Program on Key Re-search and Development Project (2016YFA0400800) and the National Natural Science Foundation of China under grants 11733009, U1838201 and U1838202. This work made use of data from the {\em Insight}-HXMT mission, a project funded by China National Space Administration (CNSA) and the Chinese Academy of Sciences (CAS).
Support for this work was provided by Universit\`a di Ferrara through grant FIR~2018 ``A Broad-band study of Cosmic Gamma-Ray Burst Prompt and Afterglow Emission" (PI Guidorzi). We acknowledge financial contribution from the agreement ASI-INAF n.2017-14-H.0.
\end{acknowledgements}




\end{document}